# Magnetization reversal in Fe(001) films grown by magnetic field assisted molecular beam epitaxy

B. Blyzniuk,[1] A. Dziwoki,[1,2] K. Freindl,[1] A. Kozioł-Rachwał,[3] E. Madej,[1,*] E. Młyńczak,[1] M. Szpytma,[3] D. Wilgocka-Ślezak,[1] J. Korecki,[1] and N. Spiridis[1]

[1]Jerzy Haber Institute of Catalysis and Surface Chemistry, Polish Academy of Sciences, Niezapominajek 8, 30-239 Krakow, Poland;

[2]PREVAC sp. z o.o., Raciborska Str. 61, 44-362 Rogów, Poland

[3]AGH University of Krakow, Faculty of Physics and Applied Computer Science, al. Mickiewicza 30, 30-059 Kraków, Poland

[*]corresponding author, ewa.madej@ikifp.edu.pl

## Highlights

- Magnetic field assisted molecular beam epitaxy was used for growing Fe(001) films on MgO(001).
- Fe(001) films deposited in an external magnetic field have distinctly different morphology from films grown with no field.
- Magnetization reversal process in Fe(001) films was studied using magneto-optic Kerr effect magnetometry and microscopy.
- Magnetization reversal that occurs via 90° domains differs for the no-field and in-field grown samples.

## Abstract

We studied the influence of a magnetic field (MF) on epitaxial growth and magnetic properties of Fe(001) films deposited on MgO(001). Thanks to modular sample holders and a specialized manipulator in our multi-chamber ultrahigh vacuum system, the films could be deposited and annealed in an in-plane MF of 100 mT. *In situ* scanning tunnelling microscopy showed that MF had a strong influence on the film morphology, and, in particular, on the structure of surface steps. The magnetic properties were studied *ex situ* using magneto-optic

Kerr effect (MOKE) magnetometry and microscopy. We showed that the moderate in-plane magnetic field applied during growth has the visible impact on the magnetic properties. The observed angular dependence of the MOKE loops and domain structures were discussed based on a magnetization reversal model. In particular we found that magnetization reversal occurs via 90° domains and the reversal differs for the no-field and in-field grown samples, in correlation with the film morphology.

**Keywords:** Magnetization reversal, Iron thin films, Magneto-optic Kerr effect, Magnetic anisotropy, Magnetic domains

## 1. Introduction

Epitaxial growth of iron on MgO(001) has a long history [1], and nowadays this system is considered as a model thin film epitaxial ferromagnet with a vast of interesting properties and applications [2]. A special attention is paid to the magnetic anisotropy and magnetization reversal processes. For thin Fe films the crystal anisotropy is low compared to the shape anisotropy. Consequently, the magnetization reversal processes are confined to the film plane and the four-fold in-plane symmetry of the magnetic anisotropy is expected for ideal Fe(001) films, with the in-plane Fe[100] and Fe[010] easy axes. Magnetization reversal observed in magneto-optic Kerr magnetometry experiments (occasionally accompanied by Kerr microscopy) on Fe(001) epitaxial films with different substrate/overlayer configurations were interpreted by a combination of coherent rotation and nucleation or/and displacement of the domain walls. Depending upon the field orientation, the reversal can proceed either via a "1-jump" mechanism, by sweeping of 180° domain walls (which gives a classic square hysteresis loop), or by a "2-jump" mechanism, by sweeping of 90° domain walls at two distinct applied field strengths, which gives a more unusual hysteresis loop with two irreversible transitions [3, 4, 5, 6]. A frequently observed special feature of the magneto-optic Kerr effect (MOKE) loops is a pronounced asymmetry that depends on the optical experimental parameter and dielectric tensor of the material [7]. Postava et al. [8] and Yan et al. [6] explained and quantified this asymmetry for Fe(001) films as arising from quadratic magneto-optical effects which cause the transverse magnetization component to contribute to the longitudinal magnetization loops.

Frequently, the interpretation of the magneto-optic data for Fe(001) films requires breaking the four-fold anisotropy symmetry by additional (usually weak) uniaxial anisotropy. Essentially, two different sources of the uniaxial anisotropy is considered: (i) low symmetry of the substrate and (ii) geometry of oblique deposition. A good example of the first case are the Fe(001) on GaAs(001) [3, 5], where it was convincingly proven that the two-fold symmetry caused by orientations of As and Ga dangling bonds breaks the four-fold symmetry of the Fe lattice at the interface [9]. The representative example of the second case are the Fe(001) films on MgO(001) [10, 11, 12, 13, 14]. Uniaxial anisotropy in the Fe(001)/MgO(001) films can be also induced by the direct modification of the film morphology by ion sputtering [15, 16].

The above cited references show that the magnetic properties of the Fe(001)/MgO(001) films are very sensitive to deposition conditions, and different growth parameters may lead to modification of the magnetic anisotropy and its symmetry. As one of such parameters an external magnetic field applied during film growth can be implemented. This growth methodology was used in the past for shaping structural and magnetic properties of thin oxide and metal films prepared via different methods from the gas phase: pulsed laser deposition [17], atomic layer deposition [18], chemical vapor deposition [19], magnetron sputtering [20] or electron beam evaporation [21]. In our recent paper [22] we showed that an external magnetic field applied during molecular beam epitaxy (MBE) growth of epitaxial magnetite films is an important agent in tailoring film morphology and, consequently, the magnetization reversal process. In the present contribution we use the methodology of magnetic field assisted MBE for growing Fe(001) films on Mg(001). We show that the morphology of a film grown in-field is distinctly different from the no-field growth, which has the direct impact on the magnetization reversal process. By combination of MOKE magnetometry and microscopy we essentially contribute to understanding of the magnetization reversal process that occurs through 90° domain walls and we show that details of the reversal differs for the no-field and in-field grown samples, in correlation with the film morphology.

## 2. Materials and methods

Fe(001) films, typically 10 nm thick, were grown on MgO(001) substrates using MBE in a multi-chamber ultra-high vacuum (UHV) system (base pressure $2*10^{-10}$ mbar). Prior to deposition of iron the substrates were annealed at 600 °C, and a 3 nm homoepitaxial MgO buffer layer was grown from an electron beam evaporator to ensure the clean carbon free MgO(001) surface. The $^{57}$Fe isotope was deposited using a home-built MBE evaporator (BeO crucibles) on the substrate kept at room temperature (RT) and eventually the films were annealed at 400 °C. During deposition the pressure remained in the $10^{-10}$ mbar range. The film thickness was controlled using a quartz crystal monitor within ±3% accuracy and ±1% reproducibility. After *in situ* characterization the samples were covered by a 3 nm epitaxial MgO(001) protecting layer for *ex situ* measurements.

A special feature of our UHV system (PREVAC) is a modular construction of the sample holders which, as combined with a two station sample manipulator, allows a magnetic field in different configuration to be applied during the film growth and post-growth processing, as described in more details in Ref. 22. The sample stations of the manipulator accept so-called "PTS" (PREVAC) and flag-style (FS) sample holders [23]. The PTS sample holders are specialized for different functions and can be operated as adapters for the FS holders carrying substrates. The range of the PTS adapters includes the holders transmitting the magnetic field from samarium–cobalt permanent magnets to a sample mounted on the FS holder. Both in-plane (up to 100 mT), used in the present study, and out-of-plane (up to 200 mT) fields are possible, the former along different in-plane direction, depending on the sample orientation on the FS holder. During the magnetic field assisted deposition the sample can be resistively heated up to 500 °C without degrading the Sm-Co magnets that maintain their working temperature thanks to a water or liquid nitrogen cooling system. At the preparation stages requiring even higher temperatures, the FS holders can be heated by electron bombardment in the FS holder station or a specialized PTS holder. Three types of samples were prepared: reference samples, deposited with no external magnetic field ($B_{depo}=0$), and samples with $B_{depo}=100$ mT applied in the film plane along the Fe[100] and Fe[110] axes.

The samples were characterized *in situ* using low energy electron diffraction (LEED) and scanning tunnelling microscopy (STM), whereas the magnetic properties were measured *ex situ* using conversion electron Mössbauer spectroscopy (CEMS) and MOKE magnetometry.

The MOKE loops were further interpreted by magnetic domain imaging using MOKE microscopy.

MOKE hysteresis loops were measured at RT using the polarization-modulation technique with a photoelastic modulator (Hinds Instruments, Hillsboro, OR, USA) in the longitudinal (L-MOKE) and (occasionally) in transversal geometry. The light from a 635 nm diode laser (Coherent, Santa Clara, CA, USA) illuminated the samples at an incident angle 26.6°. For domain imaging a Kerr microscope from evico magnetics GmbH was used.

## 3. Results and discussion

3.1 Structural and basic magnetic characterization

First, in view of the reported thickness dependence of the magnetic properties of Fe(001) films on MgO(001) [2, 8, 16, 24] we optimized the thickness of the Fe films for further detailed studies. For this purpose we prepared a sample with the stepped-thickness of 5 nm, 10 nm, 20 nm, and 30 nm. The LEED patterns in Fig. 1 a-d show that whereas for the 5 nm thickness the broadened LEED spots witness a disturbance of the long range crystal order (e.g. film discontinuity due to island growth [2]), above 5 nm thickness the films present a similar very good structural quality of the Fe(001) surface, with typical epitaxial relations: Fe[110]∥MgO[100]. The systematics in the LEED patterns is reflected in the easy MOKE loops shown in Fig. 1 e. The broader (coercive field $H_c$=2.3 mT) and rounded loop for 5 nm becomes rectangular and narrows to $H_c$=1.4 mT for the 10 nm film and remains practically unchanged for the thicker films. Therefore we concluded that the 10 nm thickness is sufficient to stabilize the structural and magnetic properties, and for further studies all films were 10 nm thick. All 10-nm films were characterised using CEMS. A typical CEMS spectrum (not shown) was dominated by a bulk-like component (approximately 97% of the spectral intensity) with the hyperfine magnetic field 33 T, the isomer shift of α-Fe and narrow lines. The remaining 3% (approximately 2 Fe ML) originates from the interface with MgO, in accordance with our previous study [25].

The key observation for the present study is the influence of $B_{depo}$ on the film morphology imaged using STM. Figs 1 f and g compare STM images for 10 nm Fe(001) films on MgO(001) deposited at $B_{depo}=0$ and $B_{depo}=100$ mT parallel to the Fe[100] easy direction, respectively. The corresponding surfaces differ in the step and terraces morphology. For the no-field deposition the structure of steps (that are actually screw dislocations) and terraces is irregular. There is no preferred step directions but the surface is flat and the RMS roughness over the presented image is only 0.078 nm. The in-field deposition resulted in a distinctly different morphology, with steps along <100> directions, numerous apparently monoatomic terraces exposed, and greater roughness of approximately 0.116 nm. Morphology of this type is typical for the kinetically limited growth, when the thermal energy is insufficient to overcome the energy barriers at steps or when the diffusion is anisotropic [26]. It seems that $B_{depo}$ parallel to Fe[100] hampers the step flow of the Fe atoms during growth or post-deposition annealing.

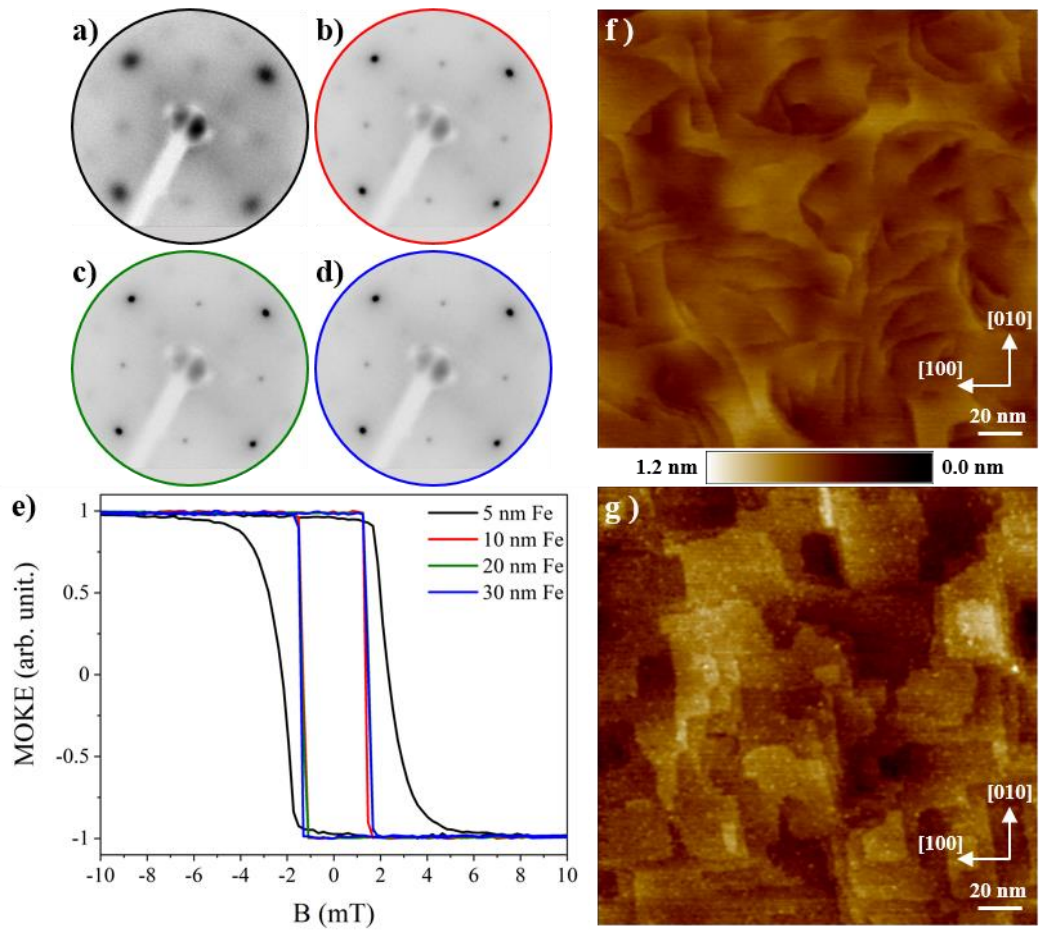

Fig. 1. (a-d): LEED images (electron energy 160 eV) for iron films with thicknesses 5 nm, 10 nm, 20 nm and 30 nm, respectively, (e): corresponding MOKE loops along Fe[100]. STM images for 10 nm Fe(001) film deposited with no magnetic field (f) and at $B_{depo}=100$ mT (g) applied parallel to Fe[100]. The height scale applies for both STM images.

3.2 Effect of in-field deposition on magnetic properties

To verify the effect of $B_{depo}$ on the magnetic properties, we measured the full dependence of the MOKE loops on an azimuthal angle θ between the direction of the magnetic field B and the in-plane crystallographic axes (θ=0º corresponds to B ∥ [100]) by rotating the sample around the normal, so that the magnet and optical parameters were unchanged. The L-MOKE loops were measured in a field range ±110 mT that is enough to saturate magnetization in the hard direction (the saturation field for the measured samples is between 50 mT and 65 mT). Exemplary L-MOKE loops measured with p-polarized light for selected θ are shown in Fig. 2 a and b for the reference sample ($B_{depo}$=0) and for $B_{depo}$ ∥ [100], respectively (the loops for the sample with $B_{depo}$ ∥ [110] are similar to the reference sample). For clarity the field range is limited to ±30 mT, which is the range of essential loop changes. The loops show typical for this system angular dependence: along the easy axes (e.g at θ=0º) the loops are rectangular with one step and, deviating to the hard in-plane axis (e.g at θ=45º), they become more complex, with two jumps that are defined in Fig. 2 b. The loops can be quantitatively analysed in a model combining the coherent rotation and the domain wall pinning/displacement, as proposed by Yan et al. [6], however, this analysis is beyond the scope of the present contribution, which aims at presenting the $B_{depo}$ effect. This effect is best seen in Figs 2 c-e, which summarize the results of the angular measurement. First, we note that there is no significant effect of $B_{depo}$ on the four-fold symmetry of the remanence magnetization (compare Fig. 2 c), which means a negligible contribution of additional uniaxial anisotropy. On the other hand, whereas $B_{depo}$ applied along the hard direction does not essentially affect the magnetization reversal process, as seen from the angular dependence of the $2^{nd}$ and $1^{st}$ switching field in Figs 2 d and e, respectively, $B_{depo}$ along the easy direction considerably enhances the coercivity of the easy loops and affects the azimuthal dependence of the reversal processes. Whereas for the reference sample (Fig. 2 a) the loops exhibit the one-jump character up to θ ≈ 25º, for the sample deposited at $B_{depo}$=100 mT parallel to Fe[100] the $2^{nd}$ jump appears already at a deviation from the easy axis as small as several degrees. What is more, the $1^{st}$ jump that reflects the anatomy of the magnetization reversal periodically changes with θ, in the opposite phase to the angular dependence of the $2^{nd}$ jumps. This indicates the decisive role of the regular substrate step morphology in nucleation and/or displacement of the domain walls.

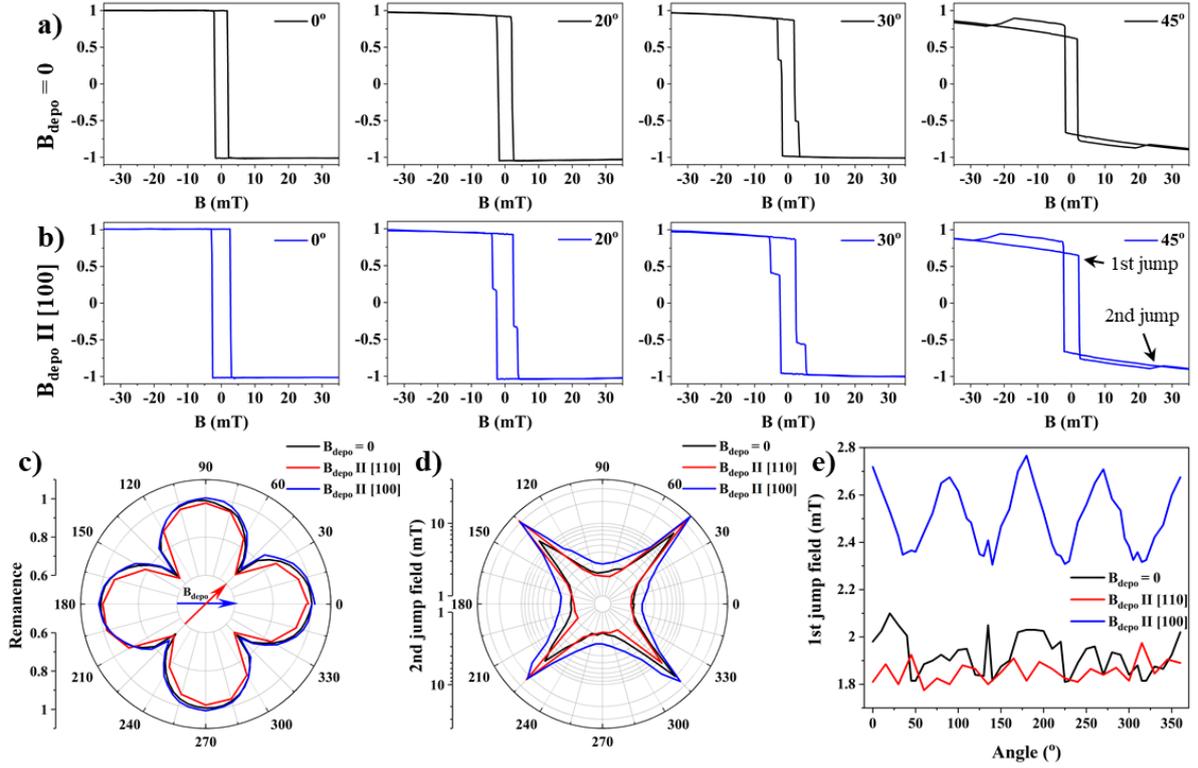

Fig. 2. Examples of Kerr loops measured at selected angles θ for the reference sample ($B_{depo}=0$) (a) and for the sample deposited at $B_{depo}=100$ mT parallel to Fe[100] (b); (c-e) azimuthal dependencies of remanence, 2nd and 1st jump fields, respectively.

A more detailed insight into the dynamics of the domain walls and the reversal mechanism was gained by Kerr microscopy imaging and local hysteresis loops extracted from the image series. For the reference sample (Fig. 3, first column) the apparently rectangular easy hysteresis loop shows up to be more complex when derived from the images acquired with a fine magnetic field step. The jump of the "rectangular loop" has two stages: the first that starts at 0.7 mT is steep and corresponds to the magnetization reversal from the [100] (black) to perpendicular [010] (grey) domain (Fig. 3 b). This stage proceeds via nucleation and unpinning of 90° domains with the domain walls close to [110]. The second stage starts at approximately 0.8 mT and covers approximately 2 mT. In this stage (Fig. 3 c) white [$\bar{1}$00]-domains nucleate and coexist with the black and grey ones, however, only 90°- domain walls are present, which is assured by local nucleation of small triangular domains at potential borders between [100]- and [$\bar{1}$00]-domains. The 90°domain walls between the grey [010]- and white [$\bar{1}$00]-domains quickly sweep over the sample, whereas the small triangular domains are visibly stronger pinned and persist to approximately 3 mT.

For the hard direction, the domain anatomy in the reference sample is similar, however, the first and second stage of the reversal process are separated by the characteristic field difference of the 1st and 2nd jump. In this case it is possible to determine the field range corresponding to the magnetization jumps, which is 1 mT and as much as 7 mT for the 1st and 2nd jump, respectively. Moreover, it becomes clear that similar domains and domain walls are behind the easy and hard loops. The role of domain unpinning is confirmed by observation of the identical domains for the easy and hard reversal (compare the long narrow domains in images b) and f) in Fig. 3.

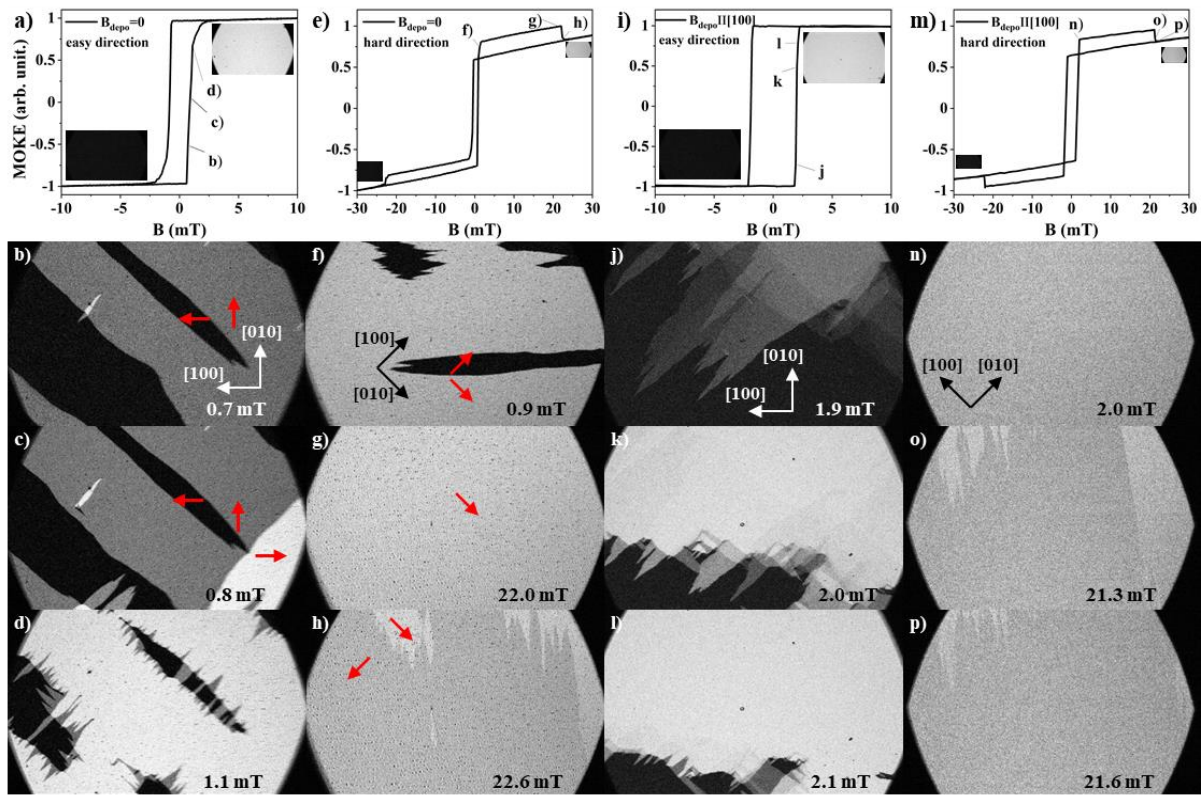

Fig. 3. Results of Kerr microscopy: local hysteresis loops and evolution of the domain structure for the reference sample ($B_{depo}=0$) along easy and hard axis, first and second column, respectively, and for the sample deposited at $B_{depo}=100$ mT parallel to Fe[100], along easy and hard axis, third and fourth column, respectively. The field of view is 500 μm, except the third column, where it is 1 mm.

A modified picture is found for the sample deposited in-field. First, along with the change in the easy L-MOKE loop that becomes broader and the steps sharper (Fig. 3 i), we observe a different domain patterns. Because the jump is very steep (the entire jump is completed across 0.4 mT), the images j) – l) in Fig. 3 result from overlapping of different domain states. Nevertheless, we notice that, albeit difficult to distinguish, the magnetization reversal along the easy axis is still a two-step process, in which the both stages of domain nucleation and propagation coexist. Similarly, also the hard loop exhibits very sharp jumps, so that no domain walls could be observed in the 1$^{st}$ jump even for a field step as small as 0.02 mT. In the same time, width of the second jump was reduced ten times compared to the reference sample, down to 0.7 mT. Our preliminary conclusion is that the regular morphology of the step structures for the in-field deposited sample is beneficial for the homogeneity of the nucleation, unpinning and propagation of the domains.

## 4. Conclusions

We showed that a moderate (100 mT) in-plane magnetic field applied during growth of epitaxial Fe(001) films on MgO(001) has the visible impact on their magnetic properties, wherein the direction of the magnetic field with respect to the film crystal orientation is decisive. Together with our recent data on magnetic field assisted growth of epitaxial magnetite films [22], to the best of our knowledge these are first experiments demonstrating the influence of the external magnetic field on the morphology and magnetic properties of metal and oxide epitaxial films. The primary reason of this effect is modification of the film morphology by application of the magnetic field during growth. For magnetic films grown under a magnetic field, the magnetoelastic effects can influence epitaxial strains leading to modification of the interface energetics and eventually the growth mode [27]. On the other hand, since the film growth proceeds far from the equilibrium conditions, these are kinetic parameters, such as diffusion rates, as well as its anisotropy and energy barriers at steps, kinks and terraces that decide on the resulting surface morphology. The dependence of these parameters on the magnetic field remains rather unexplored. The question on the origin of the observed effect - strains or kinetics - certainly deserves extended studies.

**Acknowledgments**

This research was funded by National Science Centre, Poland (NCN), grant number 2020/39/B/ST5/01838. AD acknowledges support from the Polish Ministry of Education and Science under the program "Doktorat Wdrożeniowy". AK-R and MS were supported by the program „Excellence initiative – research university" for the AGH University of Science and Technology.